\documentclass[prb,longbibliography,superscriptaddress,showpacs,aps,twocolumn]{revtex4-1}
\usepackage{graphicx}

\begin{document} 
\title{Thermodynamic evidence for the Bose glass transition in twinned YBa$_2$Cu$_3$O$_{7-\delta}$ crystals}

\author{D.J. P\'erez-Morelo}
\affiliation{Laboratorio de Bajas Temperaturas, Centro At\'omico Bariloche and Instituto Balseiro, Comisi\'on Nacional de Energ\'{\i}a At\'omica, Av. Bustillo 9500, R8402AGP S. C. de Bariloche, Argentina}
\author{E. Osquiguil
}
\affiliation{Laboratorio de Bajas Temperaturas, Centro At\'omico Bariloche and Instituto Balseiro, Comisi\'on Nacional de Energ\'{\i}a At\'omica, Av. Bustillo 9500, R8402AGP S. C. de Bariloche, Argentina}
\author{A. B. Kolton}
\affiliation{Teor\'ia de la Materia Condensada, Centro At\'omico Bariloche and Instituto Balseiro, Comisi\'on Nacional de Energ\'{\i}a At\'omica, Av. Bustillo 9500, R8402AGP S. C. de Bariloche, Argentina}
\author{G. Nieva}
\affiliation{Laboratorio de Bajas Temperaturas, Centro At\'omico Bariloche and Instituto Balseiro, Comisi\'on Nacional de Energ\'{\i}a At\'omica, Av. Bustillo 9500, R8402AGP S. C. de Bariloche, Argentina}
\author{I. W. Jung}
\affiliation{Center for Nanoscale Materials, Argonne National Laboratory, Argonne, IL 60439, USA}
\author{D. L\'opez}
\affiliation{Center for Nanoscale Materials, Argonne National Laboratory, Argonne, IL 60439, USA}
\author{H. Pastoriza}
\affiliation{Laboratorio de Bajas Temperaturas, Centro At\'omico Bariloche and Instituto Balseiro, Comisi\'on Nacional de Energ\'{\i}a At\'omica, Av. Bustillo 9500, R8402AGP S. C. de Bariloche, Argentina}
\date{
\today}

\begin{abstract}
We used a micromechanical  torsional oscillator to measure the magnetic response of a twinned YBa$_2$Cu$_3$O$_{7-\delta}$ single crystal disk near the Bose glass transition. We observe an anomaly in the temperature dependence of the magnetization consistent with the appearance of a magnetic shielding perpendicular to the correlated pinning of the twin boundaries. This effect is related to the thermodynamic transition from the vortex liquid phase to a Bose glass state.    
\end{abstract}

\pacs{74.25.Uv, 74.72.-h, 74.25.Bt, 74.25.-q}

\maketitle

\section{Introduction}
The magnetic field - temperature phase diagram of vortex matter in high temperature superconductors exhibits a variety of thermodynamic phases and phase transitions whose properties are strongly influenced by the nature of the structural defects in the superconductors.\cite{blatter94}
Particularly interesting is the case of spatially correlated defects, like twin boundaries and columnar tracks generated by heavy ion irradiation, because they produce correlated pinning on the vortex ensemble and can induce a new thermodynamic state called Bose glass.\cite{nelson92} This Bose glass phase is characterized by the collective alignment of the vortices into the correlated defects and has two distinctive characteristics:
a vanishing of the linear electrical resistivity and a diverging elastic tilt modulus.\cite{nelson93} As a result of this vortex localization, the Bose-glass phase will exhibit a transverse Meissner effect\cite{nelson93}: a finite transverse critical magnetic field is required to tilt the vortex lines away from the correlated defects. For magnetic fields larger than this critical transverse field, the vortex system acquires a tilted pinning state\cite{sonin93} where vortices are partially locked along the correlated defects. In this state, usually identified as the staircase configuration, 
the capability to shield perpendicular magnetic fields is considerably reduced. The transition from the high temperature liquid state to the Bose glass phase will only occur when the angle between the external magnetic field and the correlated defects is small enough. Theory predicts that the angular dependence of the Bose glass critical temperature has a sharp maximum when the vortices are perfectly aligned with the correlated defects.\cite{nelson93} 

Many experiments to test this phase transition have been performed in different superconductors with correlated defects and can be classified in two groups: transport measurements\cite{maiorov01,grigera98} and shielding current measurements.
\cite{smith01,zhukov97} 
From the first group of experiments  clear evidence for the transition to a Bose glass state has been obtained through the scaling of the voltage versus current data and the angular dependence of the critical temperature.  Although these measurements show critical scaling its thermodynamic properties are hindered because of the out of equilibrium nature of the experiments. 
On the other hand, magnetic measurements\cite{smith00} have shown evidence for an enhancement of the shielding capability to perpendicular magnetic fields, but again these experiments were done far from equilibrium and can not discard that the observed effects were due to some metastable configuration.\cite{smith00} In this paper we used a micromechanical torsional oscillator to study the Bose glass transition in a micron size twinned YBa$_2$Cu$_3$O$_{7-\delta}$ crystal. Our results provide direct evidence of the transverse Meissner effect characteristic of this phase. 

\section{Experimental Details} 
 In this work we use a micromechanical oscillator device similar to the one used by 
Chan {\em et al.} \cite{chan01} and Decca {\em et al.} \cite{decca03,intravaia13} to obtain precise measurements of the Casimir force. Micromechanical oscillators have been used as  magnetometers for microscopic magnetic and superconducting samples \cite{bolle99} due to their high sensitivity,\cite{dolz} fast response and good performance at very high magnetic fields.\cite{aksyuk98}
Our device was fabricated using the multiuser process PolyMUMPs from  the commercial foundry Memscap.\cite{memscap} 
It consists of a polysilicon paddle of $500\times500\times3.5 \mu\textrm{m}^3$ anchored to the substrate by two $200\mu\textrm{m}$ long parallel rods of section $2\times2\mu\textrm{m}^2$ on each extreme. The torsional elastic constant for this design is $k= 5.41 \times 10^{-9} \textrm{N}\textrm{m}\textrm{rad}^{-1}$.  Two symmetric electrodes placed on the substrate underneath each flap allows the capacitive detection of the oscillator displacement. 
In Figure 1a we show a Scanning Electron Microscope picture of the device.
\begin{figure}
\vspace*{-6.7cm}
\includegraphics[width=0.5\textwidth]{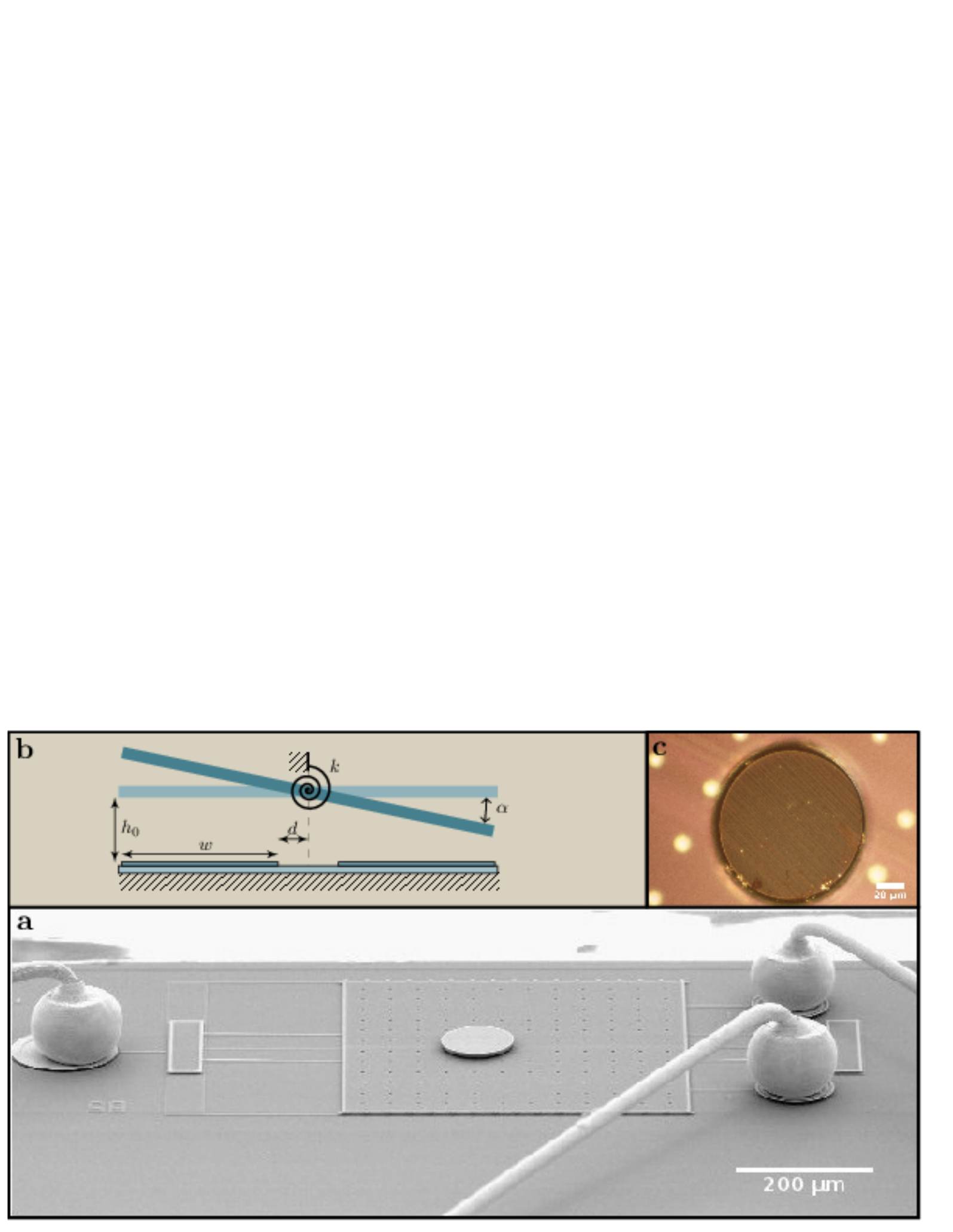}
\caption{a) Scanning Electron Microscope image of the Poly-silicon mechanical torsional oscillator with the superconducting YBa$_2$Cu$_3$O$_{7-\delta}$ (YBCO) sample mounted on it.
b) Cross section of the micro-mechanical Silicon Oscillator with the indication of the relevant geometrical dimensions. 
c) Polarized light microscopy picture of the measured disk glued to the oscillator. The normal of the twin planes were at 75 degrees from the rotational axis of the paddle.
}
\end{figure}

A circular YBa$_2$Cu$_3$O$_{7-\delta}$ (YBCO) single crystal with oriented twin boundaries is placed on top of the micromechanical oscillator (Figure 1a). The circular disk was carved out from a larger single crystal using a Focused Ion Beam milling tool. 
The disk final dimensions are $10\mu\textrm{m}$ of thickness and a diameter of $100\mu\textrm{m}$. The  sample was glued to the micromechanical oscillator with  a sub-micron drop of Apiezon N $^\copyright$ vacuum grease using glass micro-pipettes and hydraulic manipulators under an optical microscope. Ultrasonic bonding was used to wire bond 25$\mu$m gold wires to connect the oscillator to a sapphire substrate which was later mounted in a 
rotatable holder of a Oxford Teslatron $^\copyright$ variable temperature insert. The rotational axis of the holder is perpendicular to the uniform magnetic field generated by a superconducting magnet. 

All data shown in this work were taken by sweeping temperature while keeping constant the angle $\theta_H$ between the magnetic field, $\bf\vec{H}$, and the substrate.  The experiments were performed using a field cooling protocol, e.g., the data were taken while lowering the temperature after applying the magnetic field at temperatures higher than the critical one. 
When the sample become superconducting,  a magnetic torque, \begin{equation} \tau = \bf \vec{M} \times \vec{H} \label{eq1}\end{equation} appears in the sample and the torsional plate of the micromechanical oscillator will rotate around its torsional axis.\cite{decca03} This rotation is counteracted by the restoring torque caused by the deformation of the springs causing the plate to reach a tilt angle  $ \alpha= \tau / k$,
where $k$ is the oscillator's torsional spring
constant.\cite{decca03}
In our experiment, the tilt angle $\alpha$ was obtained by measuring, with an Andeen-Hagerling AH 2700A bridge, the capacitance between the oscillator paddle and each of the electrodes symmetrically placed underneath.
This capacitance is modeled by the expression
\[
C_{1,2}=\frac{\epsilon_0 L}{\tan(\pm\alpha)} \ln\left(\frac{h_0+(d+w)\tan(\pm\alpha)}{h_0+d\tan(\pm\alpha)}\right),
\] 
where $L$ is the paddle length, $h_0$ is the distance between the oscillator and the electrodes at rest,  
$d$ and $w$ are the  distances that describe the electrode geometry, and the $+$,$-$ signs correspond to the electrodes $1$ or $2$ respectively (see  Figure 1b).
  Each angular value we report was obtained from the average of the measurements from both electrodes.  

\section{Results}
\begin{figure}
\vspace*{-1cm}
\includegraphics[width=0.6\textwidth]{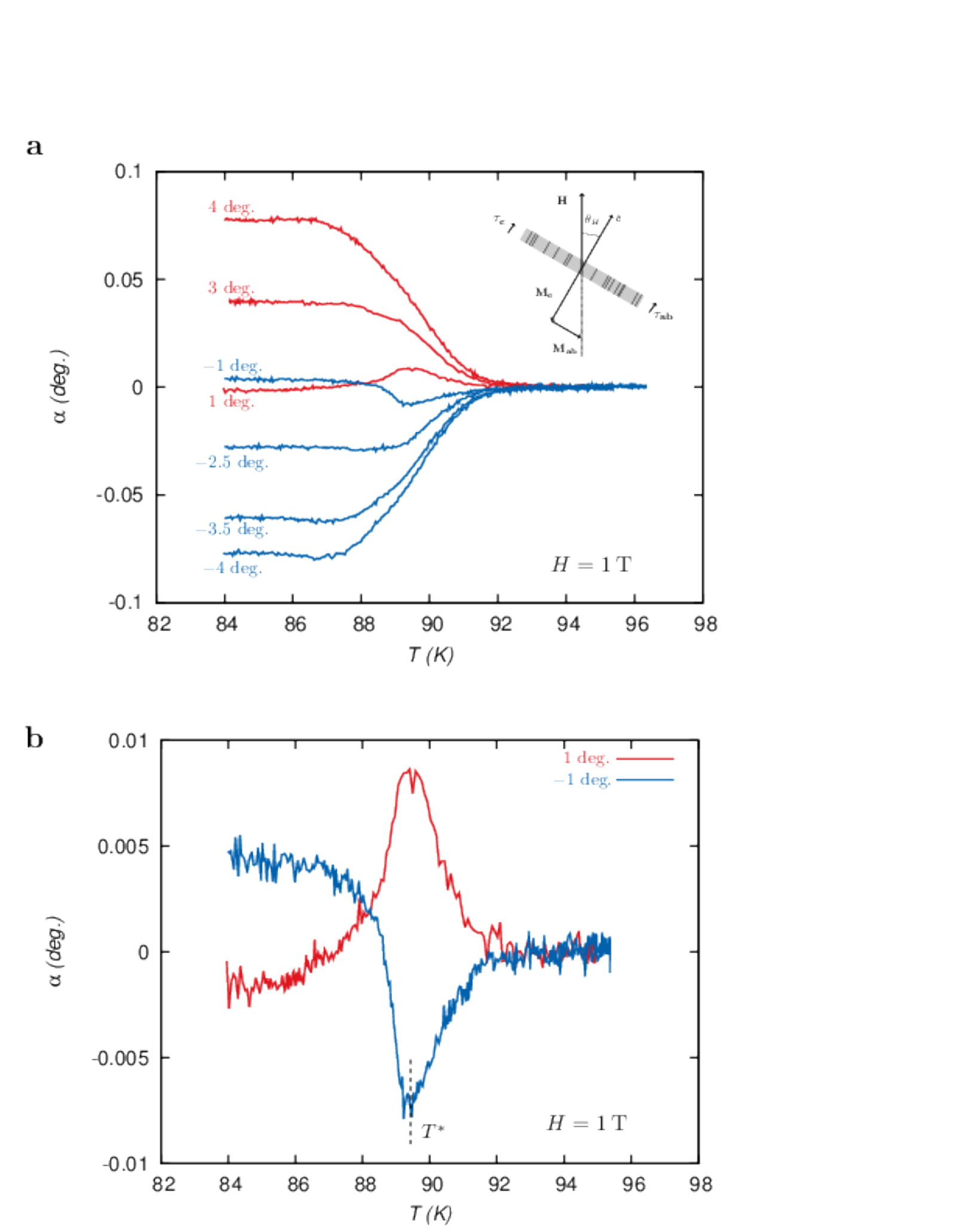}
\caption{
a) Angular displacement of the Silicon torsional oscillator, $\alpha$, as a function of temperature with an applied external magnetic field of 1T for different 
orientations, $\theta_H$, between the substrate and the magnetic field. 
All data are take from field cooling experiments, and show reversibility upon warming.
For angles $|\theta_H| < 1.5\,$deg the curves become non monotonic presenting a maximal value at a given temperature.
The inset sketch a vectorial representation of the sample magnetization indicating the different angles involved in this experiment. 
In part b)  we plot representative data for small $\theta_H$ showing the symmetric response around the c-axis direction.
}
\end{figure}

In Figure 2 we plot the oscillator angular response as a function of temperature for an applied magnetic field $\mu_0H= 1\,$T oriented at different angles close to the  c-axis direction.  We observe several features in the data: first, for all measured angles the oscillator starts to deflect at $91.8\,$K coincident with the 
superconducting critical temperature of the YBCO. The deflection is symmetric for angles, $\theta_H$, around the c-axis, and its magnitude increases as $\theta_H$ increases (in agreement with equation \ref{eq1}). This indicates that the micro-oscillator behaves essentially as a very sensitive magnetometer. 
Since the elastic constant of the torsional spring of our oscillator is temperature and field independent, for a fixed magnetization direction  the deflection angle is proportional to the sample magnetization. 
The orientation of the magnetization for a given sample is the result of two competing energies. The Zeeman energy, minimized when the magnetization and the external magnetic field are parallel, and an anisotropy energy (taking into account the geometry and structure of the sample) minimized when the magnetization is aligned to an easy axis. 
For high temperature superconducting samples, due its anisotropy and plate-like shape, the direction of the reversible magnetization of the YBCO disk is almost parallel to the c-axis.\cite{brandt98b} Therefore our measurements in the high temperature range detect the magnitude of the c-axis magnetization.

\begin{figure}
\vspace*{-1cm}
\includegraphics[width=0.6\textwidth]{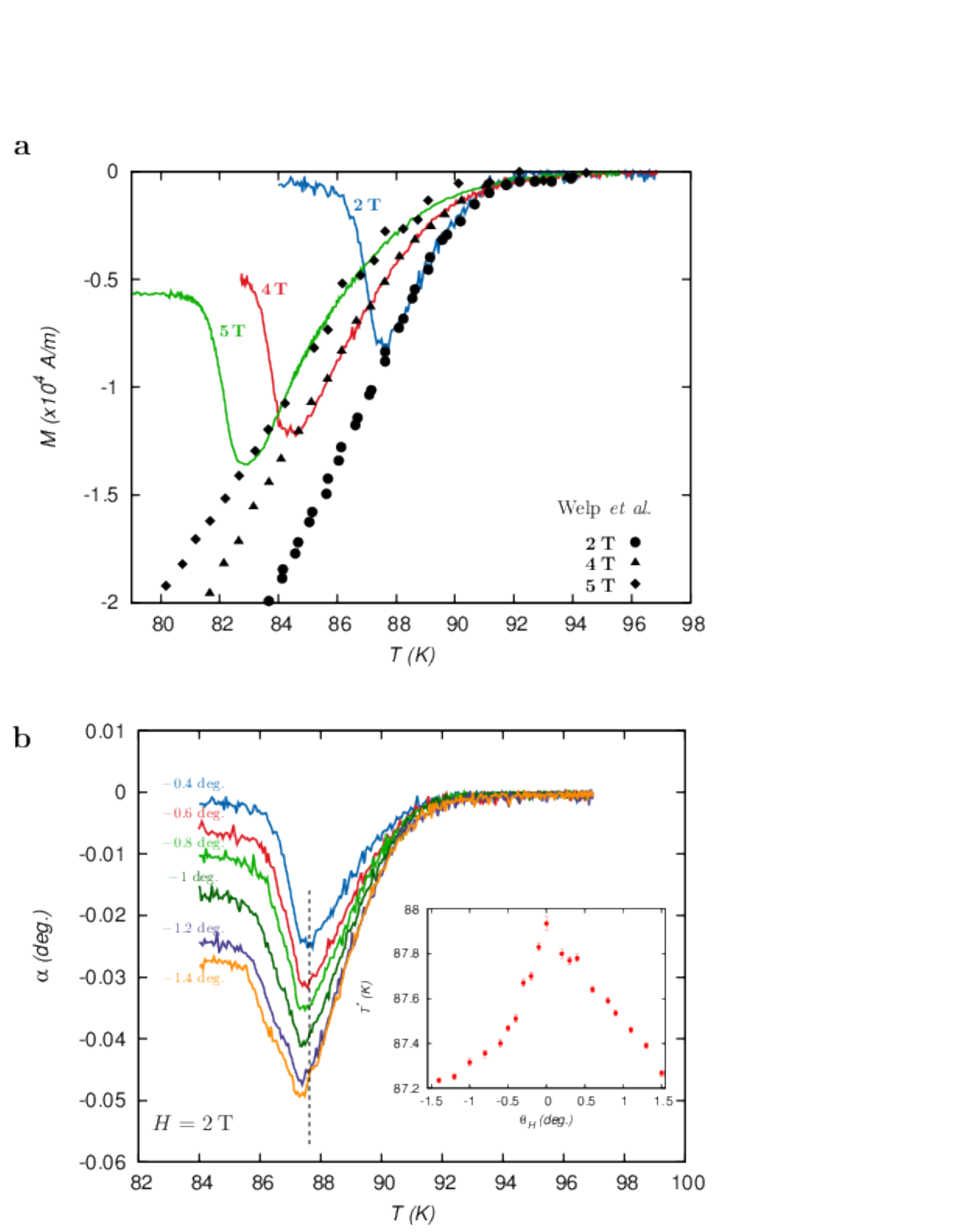}
\caption{
a) Magnetization of the superconducting YBCO disk as a function of temperature for different applied magnetic fields. Continuous lines data obtained from our experiments and discrete points values extracted from Welp {\em et al.}.\cite{welp89}
b) Angular displacements of the oscillator as a function of temperatures for an applied field of 2T for values of $\theta_H$ close to zero. A clear shift to lower values for $T^*$ as $|\theta_H|$ increases is visible.  The inset plot the dependence of $T^*$  as a function of the orientation, $\theta_H$, for an applied magnetic field of 2T.
}
\end{figure}

In Figure 3a) we plot the calculated values for the disk magnetization as a function of temperature for different magnetic fields overlaying the data from Welp {\em et al.}.\cite{welp89}

For a small range of angles around the c-axis direction we observe a clear change of the behavior: At an angular and field dependent temperature, $T^*$, the tilt angle of the micromechanical oscillator reaches a maximum  indicating that the magnetic torque exerted by the sample starts to diminish on further lowering the temperature (see Figure 2b). 
This behavior represents a sudden change in the temperature derivative of the magnetization, $\frac{d\bf M}{dT}$, which is a thermodynamic signature 
for a second order phase transition.\cite{narayan05}

A simple explanation for the measured behaviour can be given in terms of the Bose glass model.
At high temperatures, where the vortex ensemble is in the liquid state, the sample magnetization mainly points along the c-axis of the crystals due to sample geometry and material anisotropy. 
 This magnetization exerts a ``negative'' torque on the sample-oscillator system. As soon as the sample enters into the Bose glass phase a transverse Meissner effect appears. The appearance of this effect implies that the sample magnetization has an extra component that is $\approx90$ degrees away from the c-axis. Such magnetization component counteracts the torque generated by the c-axis magnetization as is schematically shown in the inset of Figure 2a. This behavior is what our micromechanical magnetometer is reflecting, a reduction of the total torque exerted by the sample.

Further evidence that this observed anomaly is related to the Bose glass transition can be obtained by plotting $T^*$ as a function of the sample angle, $\theta_H$, shown in the inset of Figure 3b, for an applied field of 2T. The cusp--like shape of the transition temperature predicted for this phase transition is very clear.
The data at this particular field shows an asymmetry between negative and positive angles. We can rule out that this effect is due to an experimental artifact like a magnetic field or temperature gradient at the sample position, an asymmetric response of the oscillator, or a thermal decoupling between the thermometer and the sample.
One possibility for this asymmetric condensation energy of the Bose glass phase is a non uniformity at the samples edges. In micron sized samples like the reported here there is a significant fraction of vortices that are affected by surface effects. We must also remark that in a previous work\cite{maiorov01}  was observed a change in the $T_{BG}$ vs $\theta$ dependence as a function of the applied magnetic field. At low fields a linear dependence was reported and at high fields a sharper cusp--like functionality was found. In our case we are measuring both dependences at the same applied magnetic field but 
at negative or positives angles.

\begin{figure}
\vspace*{-5.5cm}
\includegraphics[width=0.5\textwidth]{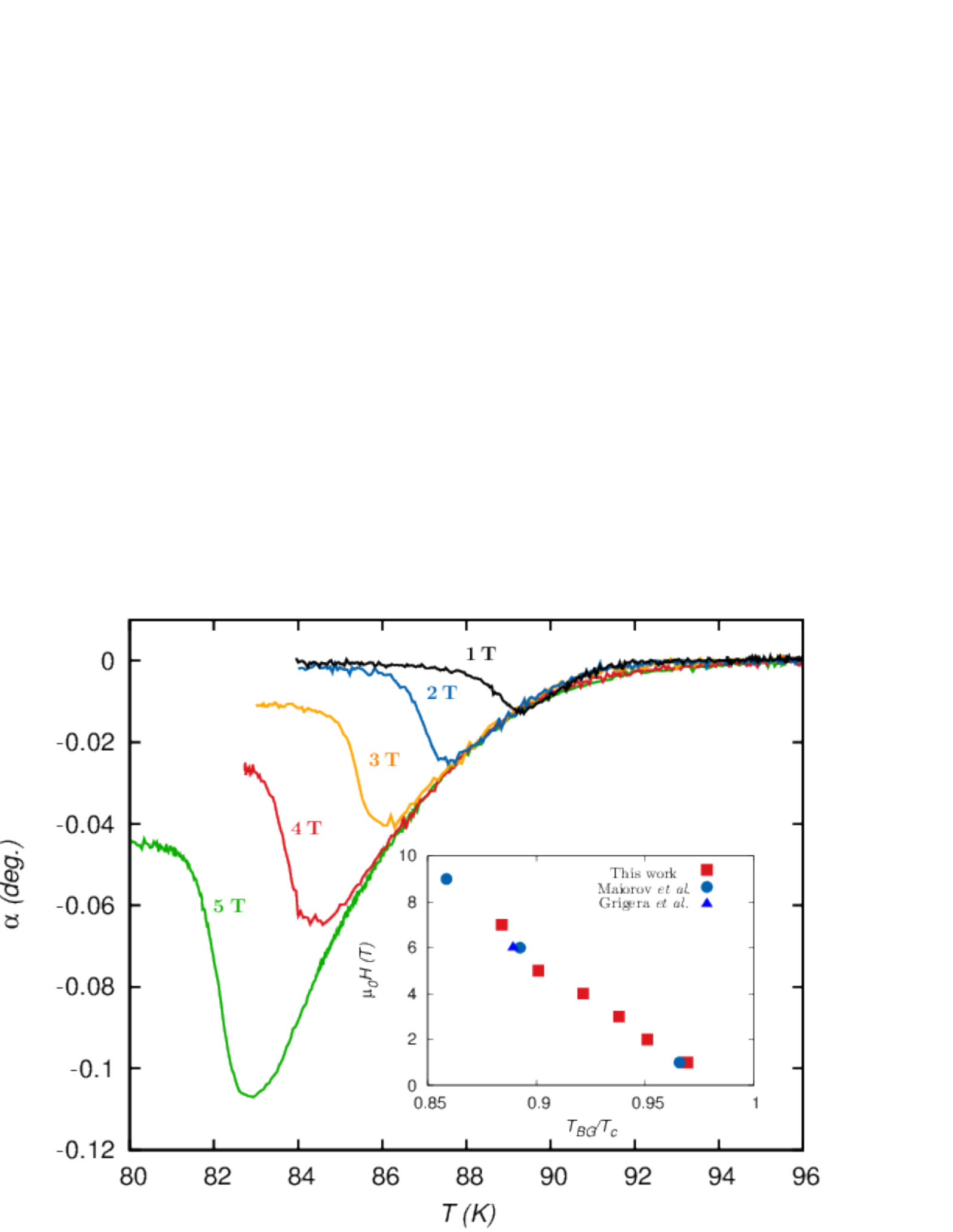}
\caption{a) Displacement angle of the oscillator as a function of temperature for different magnetic fields applied at an angle $\theta_H$=-0.4 deg. 
Inset: Magnetic field versus Temperature phase diagram for the measured $dM/dT$ discontinuity in comparison with Bose glass critical temperatures obtained from transport measurements reported by Grigera {\em et al.},\cite{grigera98} and Maiorov {\em et al.}.\cite{maiorov01} 
}
\end{figure}

In the inset of Figure 4 we summarize the position of the measured anomaly in a H-T phase diagram. For comparison, data obtained from the scaling of transport measurements\cite{grigera98,maiorov01} for similar crystals are plot in the same graph, showing a perfect agreement with our data.

Figure 4 shows representative data taken at the same nominal $\theta_H$ at different magnetic fields. A clear field dependence for the $\frac{d\bf M}{dT} $ discontinuity is observable. Moreover, the curves for all fields merge for temperatures above this feature indicating that the magnetization has a close to $1/H$ dependence in the reversible vortex liquid state. 
At temperatures closer to the critical one this data overlap is lost, indicating that a different scaling functionality must be taken into account. In this temperature range 3D thermal fluctuations on the thermodynamic magnetization were observed.\cite{welp91} 

\section{Conclusions}
In summary we have presented experimental evidence of the thermodynamic nature of the vortex system Bose glass transition in high quality YBa$_2$Cu$_3$O$_{7-\delta}$ single crystals. The temperature and angular dependences of the magnetic response, measured by a
high sensitive Silicon micro oscillator is fully consistent with a continuous transition at $T_{BG}$ and correspond to the appearance of a spontaneous magnetic shielding  to the perpendicular component of the magnetic field.

\begin{acknowledgments}DJPM Fellowship holder of Consejo Nacional de Investigaciones Cient\'{\i}ficas y T\'{e}cnicas (CONICET). EO ABK, GN and HP researchers of CONICET. This work was partially supported by PIP 1122008010111001, CONICET. This work was performed, in part, at the Center for Nanoscale Materials, a U.S. Department of Energy, Office of Science, Office of Basic Energy Sciences User Facility under Contract No. DE-AC02-06CH11357. We thank V. Bekeris for a careful reading of the manuscript and valuable suggestions. 

\end{acknowledgments}
\bibliography{meissnert}

\end{document}